\documentclass[11pt]{article}
\usepackage{fullpage}

\usepackage{graphicx}
\usepackage{url}
\usepackage{comment}
\usepackage{fullpage}

\usepackage[noend]{algorithmic}
\usepackage{algorithm}

\title{Temporal Debugging using URDB}
\author{Ana Maria Visan$^*$, Artem Polyakov$^\dag$, Praveen S. Solanki$^*$, \\
\medskip
	Kapil Arya$^*$, Tyler Denniston$^*$ and Gene Cooperman$^*$ \\
\medskip
 { \hfill
 \parbox[t]{1.5truein}{\hbox{\ }}
 $^*$\parbox[t]{2truein}{
 College of Computer \\
 \hbox{\ \ \ \ } and Information Science \\
 Northeastern University \\
 Boston, MA 02115
 }
 \hfill
 $^\dag$\parbox[t]{3truein}{
 Siberian State University \\
 \hbox{\ \ \ \ }  of Telecommunications and Informatics \\
 Novosibirsk, Russia
 }
 \hfill
 } \\
 \medskip
 e-mail:  amvisan@ccs.neu.edu, artpol84@gmail.com,
	 \{psolanki,kapil,tyler,gene\}@ccs.neu.edu
}

\date{}

\begin{document}

\maketitle

\begin{abstract}

A new style of temporal debugging is proposed.  The new URDB debugger can
employ such techniques as temporal search for finding an underlying fault
that is causing a bug.  This improves on the standard iterative debugging
style, which iteratively re-executes a program under debugger control in
the search for the underlying fault.  URDB acts as a meta-debugger, with
current support for four widely used debuggers: gdb, MATLAB, python,
and perl.  Support for a new debugger can be added in a few hours.
Among its points of novelty are: (i)~the first reversible debuggers
for MATLAB, python, and perl; (ii)~support for today's multi-core
architectures; (iii)~reversible debugging of multi-process and distributed
computations; and (iv)~temporal search on changes in program expressions.
URDB gains its reversibility and temporal abilities through the fast
checkpoint-restart capability of DMTCP (Distributed MultiThreaded
CheckPointing).  The recently enhanced DMTCP also adds ptrace support,
enabling one to freeze, migrate, and replicate debugging sessions.

\end{abstract}

\section{Introduction}
\label{sec:introduction}

Programmers have long struggled under the curse of temporality in
debugging complex code.  An unknown program fault corrupts the logic
of the program, but the resulting bug may be exposed only much later
in that program.  Debuggers assist one in analyzing the logic near the
error, but only in order to build a hypothesis that might explain the
original cause or fault of the bug.  Upon producing a hypothesis, the
programmer again executes the program under debugger control with the
hope of obtaining more conclusive proof that the hypothesis is
correct.

This style of {\em iterative debugging} is only a little better than
print-style debugging.  A better solution would be a style of {\em
  temporal debugging}.  The debugger would employ {\em temporal
  search} in order to search both backwards and forwards over the
process lifetime in order to verify or falsify the hypothesis about
the fault.  The entire process lifetime is made available for analysis
by the debugger.

Such a valuable {\em temporal search} technique is outside the current
state of the art.  There are several reasons for this.  First,
many real-world programs have multiple threads running on multiple
cores, that spawn child processes or even an entire distributed
computation.  Hence, temporally debugging a single thread would
mean taking all of the threads and processes of the rest of the world
backwards and forwards through time alongside that single thread.

A second reason for the lack of temporal search is that the program
fault and error are often widely separated in time.  While a number of
reversible debuggers have been developed over the last 40~years (see
Section~\ref{sec:relatedWork} on related work), all of these reversible
debuggers have been based on some form either of logging of instructions
or of events for later replay, or of both.  Such logging requires large
amounts of storage, and so reversible debugging has until now usually
been limited to a span of time short enough so that the debugger log
does not overflow memory.

A third and more recent reason for the lack of temporal search
is that we now live in a multi-core world in which efficient programs
typically employ multiple threads.  Hence, program execution
is no longer determined by a single sequential instruction stream,
but instead by multiple instruction streams corresponding to
multiple threads.  Hence, any straightforward logging-based approach to
multi-core computing is doomed to be extreme inefficiency,
due to the high overhead of sequentializing the multiple instruction
streams.

Finally, a fourth reason is that it is no longer true that most
computations run as single processes.  It is common to
use a scripting language that frequently spawns other processes.
Both parallel and distributed computation have become widespread.
Problem-solving environments employ specialized processes that
communicate among each other.  High-level interactive technical languages
such as MATLAB employ helper processes.  Hence, moving a single
process backwards in time may require moving backwards in time
the associated mini-world of that process.

We present a new temporal debugger, URDB, that enables such technologies
as temporal search, above.  URDB stands for {\em Universal Reversible
Debugger}.  It is freely available as open source software~\cite{URDB09}.
URDB supports temporal debugging both on multi-threaded SMP architectures
and on multi-process architectures over distributed nodes.  URDB is built
on top of existing debuggers.  It currently has four {\em personality}
modules, which support:  gdb; MATLAB; python (pdb); and perl (perl -d).
It runs as a monitor on top of the debugger, and requires no modification
of the debugger itself.

URDB gains its temporal feature through the use of DMTCP (Distributed
MultiThreaded CheckPointing), a fast checkpoint-restart package.  Because
URDB is based on checkpoints (temporal snapshots of the computation)
that are saved on disk, URDB does not require any logs.  Hence, it is
practical to use URDB to debug computations that may continue for days
or even weeks.

URDB is implemented as a ``wrapper'' around the target debugger.
The wrapper consists of a relatively thin {\em debug monitor}
to implement a simple algorithm for reversible execution (see
Section~\ref{sec:algorithm}).  No source code for the target debugger
is required, and the target debugger is not modified.  For example,
the reversible MATLAB debugger based on URDB was developed without
access to the MATLAB source code.  The checkpoint-restart software and
the debug monitor are written and debugged once, and then used in many
reversible debuggers.

Current {\em URDB personalities} (target debuggers) include gdb, MATLAB,
python (pdb) and perl (perl~-d).  A personality for a new debugger can
be developed in a few hours by modifying a 250-line python template.
As a demonstration of this, URDB was used to add reversibility to the
classic curses-based BSD UNIX games of rogue and robots.  This generality
is expected to be useful in interactive simulations.  One can go back
in time and search for those user inputs that optimize the outputs.

The URDB process-level approach uses no logging, unlike the virtual
machine approach or the reverse execution approach of gdb-7.0.  Hence,
URDB, and checkpoint/re-execute in general, lacks the deterministic replay
feature of the other two approaches.  However, record/reverse-execute and
record/replay are not able to support multiple threads in their current
incarnations.  In contrast, checkpoint/re-execute, as exemplified
by URDB, can support multiple threads, multiple cores and multiple
processes --- even over the multiple nodes of a distributed computation.
To the authors' knowledge, no previous reversible debugger has been
able to support a multi-core SMP architecture.  Furthermore, 
determinism can potentially be added back in through several orthogonal
techniques described in Section~\ref{sec:relatedWork} on related work,
such as~\cite{KendoMultithreading09,DMP09,ParkEtAl09}.

In effect, we are viewing checkpoints as the ``pearls of time''
in a process lifetime.  Where race conditions may alter events
upon re-execution, re-executing past the previous checkpoint may
lead to a slightly altered, or ``parallel worlds''.  The parallel
worlds can be tied more closely to their antecedents
by logging,
by orthogonal techniques for enforcing determinism,
or by inserting additional checkpoints to provide a tighter thread
with which to string together the pearls of time.

The temporal search abilities of URDB are demonstrated by the
implementation of {\em reverse expression
watchpoints}: for a given expression, finding a previous statement and a
point in the process lifetime such that the expression value was
different from the current value, but will change to the current value
upon executing the immediately following statement.
The temporal search uses a binary search algorithm.
Instead of evaluating an expression  $n$~times over $n$~statements
as with a traditional debugger, URDB requires only
$\log_2 n$ evaluations.

For example,
if the user implements a linked list of bounded length, and if the
user program aborts with the report {\tt linked\_list\_len(X) > N},
then in a normal debugging process, the user would add frequent assert
statements to detect when that expression becomes true.  But frequent
calls to {\tt linked\_list\_len()} is computationally expensive.  So in URDB,
the user executes a reverse expression watchpoint on the expression {\tt
linked\_list\_len(X) < N}.  to discover the last instruction for which the
consistency condition was still true.  This type of approach can
be especially valuable for such notoriously error-prone tasks
as writing a garbage collector.

A further novelty of URDB is the ability to reversible debug
distributed computations by synchronously rolling back all participating
processes of a distributed computation.  All communication through
sockets and other mechanisms is also consistently rolled back.
Again, as is the case of traditional debuggers, users must be
aware of issues of non-determinism when running distributed
computations.

While not available in the current implementation, support for
OpenMPI is planned for the near future.  This will
provide a demonstration of the more general ability of DMTCP
to support distributed computations
(see Section~\ref{sec:distributedComputation}).  The experimental
section of this paper does provide some more limited
examples of debugging multiple processes within a single computer.

Section~\ref{sec:relatedWork} presents a brief history of reversible
debugging.  Section~\ref{sec:dmtcp} provides background information on
DMTCP.  Section~\ref{sec:algorithm} describes the reversibility
algorithms of the URDB monitor.  Temporal search, and in particular,
reverse expression watchpoints, are described in
Section~\ref{sec:binarySearch}.  Reversible debugging of distributed
computations is described in Section~\ref{sec:distributedComputation}
(although the implementation for that work is still in progress).
Section~\ref{sec:monitorImplementation} describes the URDB
implementation.  Finally, Section~\ref{sec:experiment} presents the
experimental results followed by the conclusion in
Section~\ref{sec:conclusion}.

\subsection{Related Work}
\label{sec:relatedWork}

{\em Reversible debuggers} (sometimes called {\em
time-traveling debuggers}) have existed at least since the
work of Grishman in 1970~\cite{Grishman70} and Zelkowitz in
1973~\cite{Zelkowitz73}.
These debuggers are based on logging and reverse
execution of individual instructions:  {\em record/reverse-execute}.
This first approach was further developed
through such landmarks as the reversible debugger for Standard~ML by Tolmach
and Appel~\cite{TolmachAppel90,TolmachAppel95},
and the recent reversible execution feature
of gdb-7.0 (released in October, 2009).

About five years ago, a second approach, {\em record/replay}
was implemented with virtual machines using record/replay
technology.  This work was based on virtual machine snapshots
and event logging.  It was first demonstrated in the work of
King \hbox{et al.}~\cite{KingDunlapChen05}, followed by VMware's Lewis
\hbox{et~al.}~\cite{VMwareReplayDebug08}, and still other examples.

Both record/replay and record/reverse-execute emphasize the advantages
of deterministic
replay.  Both approaches have difficulties in supporting SMP multi-core
architectures, where logging a serialization of synchronous instructions
appears to be difficult.

\paragraph{1. Record/reverse-execute:}
Instruction logging records the state as each instruction is executed.
The logging contains sufficient information to `` undo'' an instruction.
In addition to logging instructions, one can log external I/O,
signals, and other events, for better determinism.  Whhile the benefits
are clear, there are also significant disadvantages.  The need
for logging prevents a debugger from executing code at full native
speed.  Further, the size of the log files can also be significant.
Finally, sequential logging of multiple threads is difficult without
operating system support.

An early example of such a reversible debugger was the AIDS debugger
built by Grishman~\cite{Grishman70} in 1970.  It was used to debug
FORTRAN and assembly language on the Control Data~6600 mainframe
supercomputer. A similar example was the work of
Zelkowitz~\cite{Zelkowitz73} in 1973.
A notable continuing success using this
approach is the work of Appel and Tolmach~\cite{TolmachAppel90,TolmachAppel95},
on a reversible debugger for Standard~ML.

In order to more quantitatively measure the pros and cons of instruction
logging, we tested gdb-7.0, which was released on Oct.~7, 2009.
It includes a reversible debugging capability based on instruction
logging~\cite{Gdb09}.

Gdb-7.0 presents an excellent interactive user feel, whether
single-stepping or reverse-stepping through a C~program.  Nevertheless,
one presumes that gdb-7.0 reversible debugging was designed for
relatively short runs that would normally execute over seconds or
minutes.

Specifically, our experiments show that gdb-7.0, when running
in the forward direction with logging ({\tt target record}), gdb was
measured on a simple C~program for hashing at 96,873 times slower than
the corresponding speed of gdb-6.8.  The average memory consumption
per C~statement for logging was measured at 104 bytes per program statement.
Further, as expected, the instruction logging in gdb-7.0 did not support
a multi-threaded program.

\paragraph{2. Record/replay:}
This approach is traditionally implemented through virtual machines.
Snapshots record the state of the machines at given
intervals.  A reproducible clock is achieved through values
of certain CPU registers, such as
the number of loads and stores since startup.  This allows
asynchronous events
to be replayed according to the time of the original clock when
they occurred.
Since 2005, at least two virtual machine-based
reversible debuggers have been developed:
 for gdb~\cite{KingDunlapChen05} and for
Visual Studio with C/C++~\cite{VMwareReplayDebug08}.   
Snapshots have the huge advantage that forward execution
can be extremely fast, close to full native speed --- depending on the
frequency and overhead of snapshots, and the overhead of event logging.
However, event logging is an important requirement for virtual
machine approaches, because they depend on sensitive timing at the
instruction level.  If external events are replayed differently
from how they were originally recorded, timing is thrown off.
In particular, the interval timer used by operating systems
for context switching is thrown off, and processes are
context switched at the wrong time, with a resulting cascade
of errors as processes interact with each other at the
wrong times.

Additionally, the cost of virtual machine snapshots is high.
They typically require about 100~MB of disk space
and 30~seconds, as opposed
to a typical 10~MB and 2~seconds for DMTCP to checkpoint a gdb session.
(The cost of DMTCP checkpoints is expected to become still cheaper with
the planned introduction of incremental checkpoints.)

\paragraph{3. Checkpoint/re-execute:}
The natural level of interaction for this approach is at the
process level:  intermediate between the instruction level of
record/reverse-execute and the operating system level of record/replay.
URDB employs checkpoint files on disk.  However, an alternative
strategy would be {\em live checkpoints}.
In this approach, checkpoints are created by forking of a child
process of the debugged application.  Process-based checkpointing is
limited by the number of processes that one can simultaneously
maintain.  
{\em Gdb-6.8} and Srinivasan et al.~\cite{Srinivasan04} have
implemented {\em live checkpoints}.  These can be used to execute a
limited form of a reverse-continue command (restore previous
checkpoint, and forward continue until next to last breakpoint).
Further the ocamldebugger~\cite[Part~III, Chapter~16]{ocaml08} for
Objective Caml combines live checkpointing with the method of Appel and
Tolmach~\cite{TolmachAppel95} for reversible debugging of Standard~ML.

\paragraph{Determinism and debugging}

The DMTCP-based approach of process checkpoints does not
currently employ determinism (although this could be added using
complementary software).  Instead, DMTCP checkpoints and restarts all
processes and threads of a computation.
For example, in a distributed MPI computation,
reverse execution will quite literally reverse the execution of the
processes of the multiple hosts in a synchronous manner.
Hence, successive executions from a checkpoint may return
different results.

The issue of determinism is a particularly sticky one when running on
SMP multi-core hardware.  A logging-based approach to reversible
debugging (e.g. record/replay or record/reverse-execute)
is difficult due to the
ordering of concurrent instructions on distinct processors.  This can alter
the result of a computation.

Although not implemented in this current work, there are potential
approaches to orthogonally add determinism to URDB while running on a
multi-core architecture.  Two examples of adding determinism to
multi-core architectures are Kendo~\cite{KendoMultithreading09} and
DMP~\cite{DMP09}.  A method for adding only partial determinism is
described for PRES~\cite{ParkEtAl09}.  The PRES technique of using
Feedback generation from previous replay attempts is especially
interesting for its synergy with the URDB reversible debugger, since a
generalization of URDB would allow URDB to run PRES on an application,
while giving PRES program control with which to direct URDB when to
create checkpoints, and when to repeatedly re-execute from a given
checkpoint.  Finally, logging of I/O and certain other events can also
be added through wrappers around system calls.

\section{Background:  DMTCP Software}
\label{sec:dmtcp}

A technical overview of the DMTCP software is presented.
Further details are in~\cite{AnselEtAl09}.
DMTCP is free software distributed from \url{http://dmtcp.sourceforge.net}.
It was developed over
five years~\cite{AnselEtAl09,CoopermanAnselMa06,RiekerAnselCooperman06},
and as of this writing, it is experiencing approximately 100~downloads per
month.
Dynamic libraries are saved and restored as part of the user-space memory.
A target application can autonomously request its own checkpoint with the
{\em dmtcpaware} library API.

DMTCP employs a centralized coordinator, to which each process
of a distributed computation connects.
The centralized DMTCP coordinator is {\em stateless}.  If the
coordinator process dies, then one kills the other processes of the
computation and starts a new DMTCP coordinator.  Processes are
restarted from the {\em checkpoint images}:
{\tt dmtcp\_restart ckpt\_*.dmtcp}.  {\em Process migration}
is accomplished by moving the checkpoint images to new computer nodes.

\begin{figure*}[htb]
\begin{center}
\resizebox{!}{2.0in}
{\includegraphics{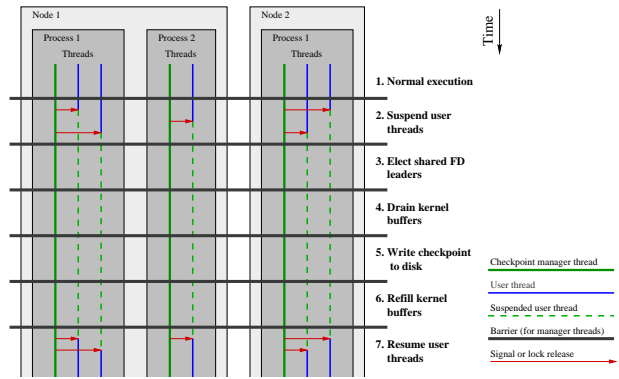}}
\end{center}
\caption{\label{fig:ckptdiag}Steps for checkpointing a simple system
with 2 nodes, 3 processes, and 5 threads.}
\end{figure*}

\begin{figure*}[htb]
  \begin{center}
  \resizebox{!}{2.0in}
  {\includegraphics{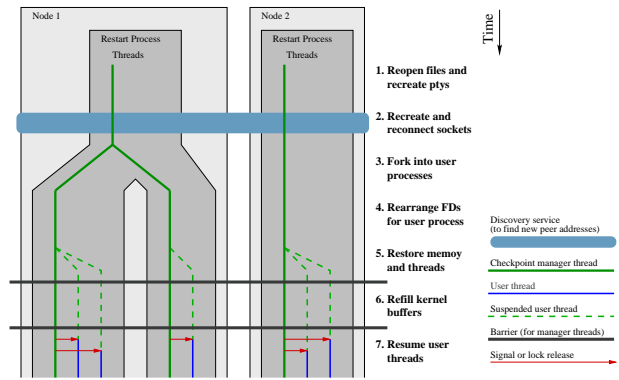}}
  \end{center}
  \caption{\label{fig:rstrdiag}
  Steps for restarting the system checkpointed in Figure~\ref{fig:ckptdiag}.
  The unified restart process and subsequent fork are required to recreate
  sockets and pipes shared between processes.
  }
\end{figure*}

Figures~\ref{fig:ckptdiag} and~\ref{fig:rstrdiag} pictorially describe
the stages of checkpointing and restarting.
Three basic requirements of distributed, user-space checkpointing
are accomplished:
\begin{enumerate}
  \item restoring user space memory;
  \item restoring kernel status; and
  \item restoring network data in flight.
\end{enumerate}
The solid horizontal lines refer to distributed barriers.
The centralized server is needed:
to enforce the distributed barriers; and
to act as a nameserver
where end-user processes register themselves
upon restart, in order to discover their peers for
reconnecting socket connections.

Among the features automatically accounted for by DMTCP are:
\begin{itemize}
\item
fork, exec, ssh, mutexes/semaphores, TCP/IP
sockets, UNIX domain sockets, pipes, ptys (pseudo-terminals), terminal modes,
ownership of controlling terminals, signal handlers, open file descriptors,
shared open file descriptors, I/O (including the readline library), shared
memory (via mmap), parent-child process relationships, pid 
and thread id virtualization.
\end{itemize}

The technical implementation uses {\tt LD\_PRELOAD} to preload a DMTCP
library, dmtcphijack.so, into each application process.  The preloaded
library creates an additional checkpoint thread for each application
process.  The checkpoint thread creates a connection to the
centralized coordinator and listens for commands.  Signals and a
signal handler are used for the checkpoint thread to capture the
``attention'' of the end-user threads.  Wrappers around the fork and
exec system calls are used to ensure that dmtcphijack.so is preloaded
into child processes.  A wrapper around exec captures any calls
to ``ssh''.  Upon {\tt ssh exec $\ldots$}, it preloads dmtcphijack.so
into the new, remote processes.  DMTCP implements wrappers for
the dynamic library {\em libc.so}.  Further implementation details
are in~\cite{AnselEtAl09}.

\section{The Debugger Monitor}
\label{sec:monitor}

\begin{figure*}[htb]
  \begin{center}
  \resizebox{!}{2.0in}
  {\includegraphics{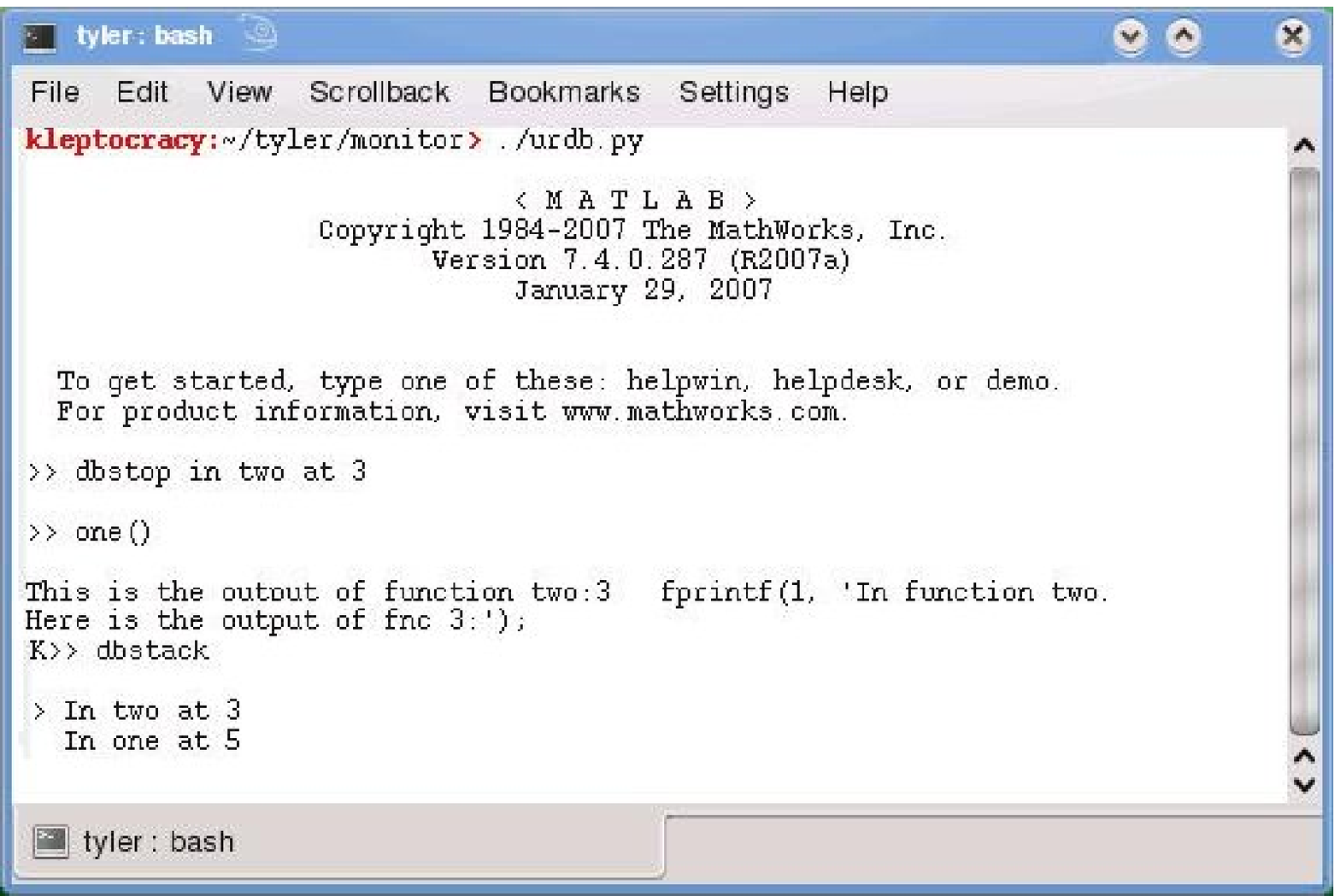}}
  \resizebox{!}{2.0in}
  {\includegraphics{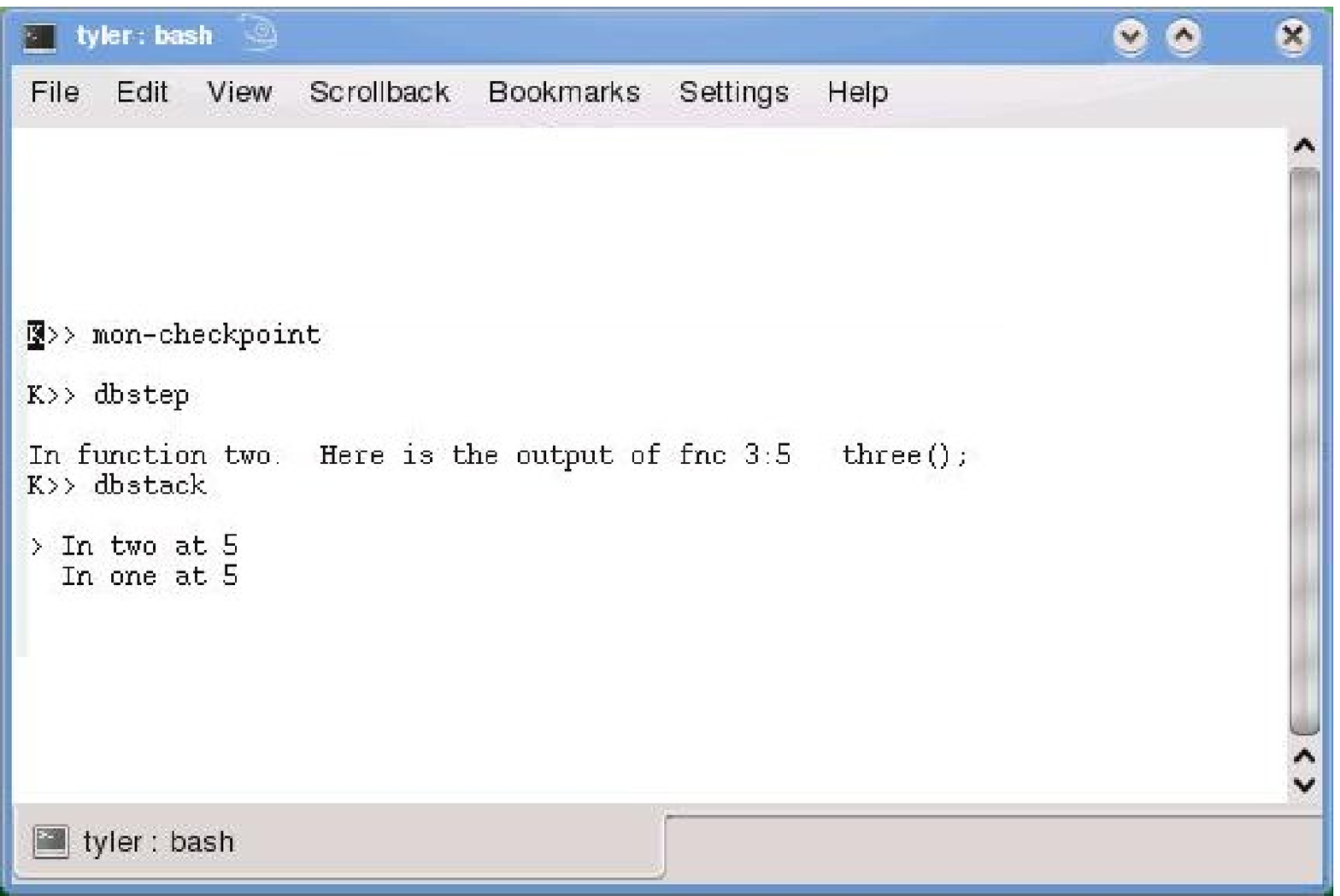}}
  \resizebox{!}{2.0in}
  {\includegraphics{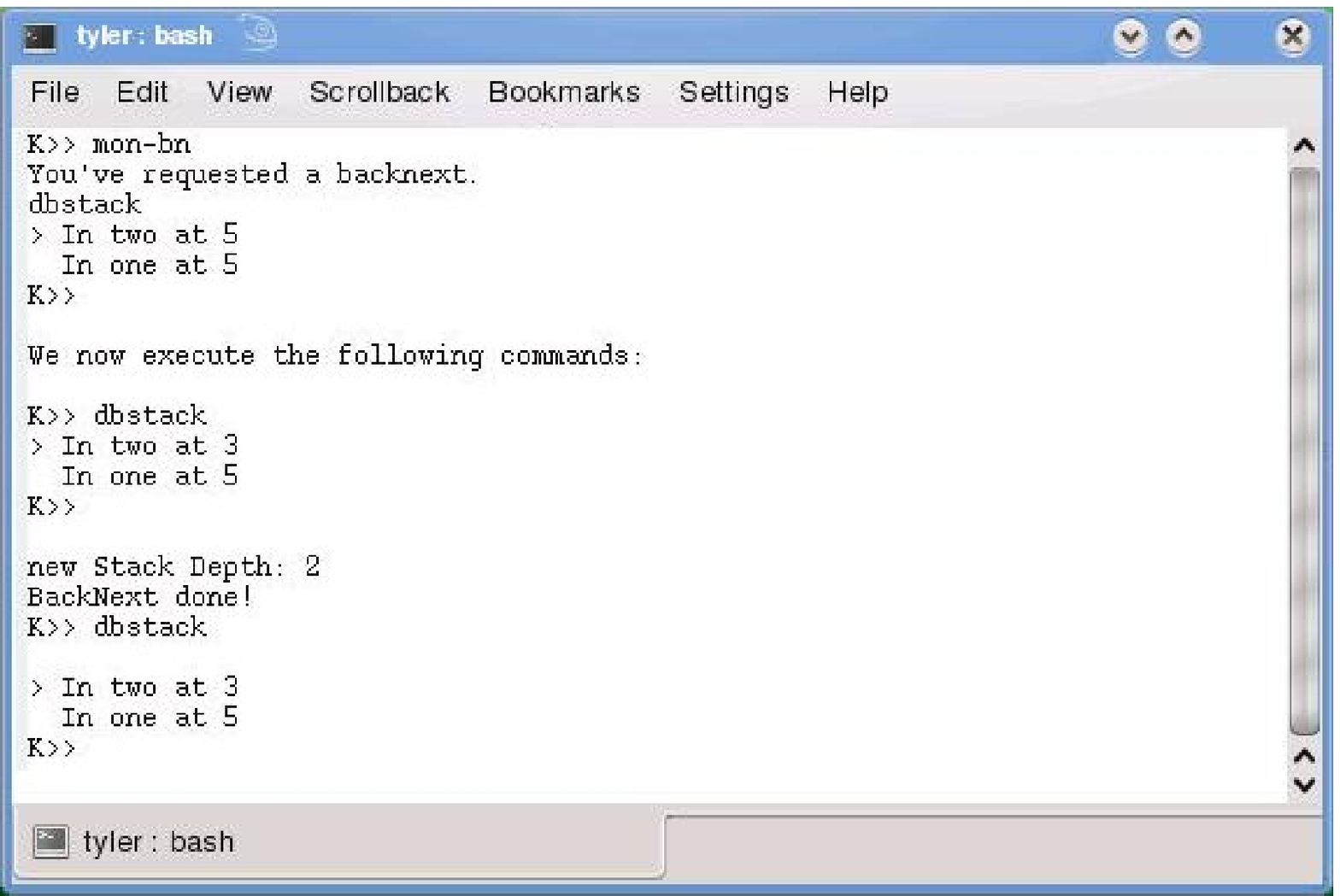}}
  \end{center}
  \caption{\label{fig:matlab}
  Reversibly debugging MATLAB code using URDB}
\end{figure*}

The URDB package consists of the DMTCP checkpoint-restart package and
a python-based debug monitor.  This prototype supports the four
debuggers: MATLAB, python, perl and gdb.  Additional debuggers can be
supported by filling in a python-based template of about 250~lines
with the names of debugger commands and regular expressions to capture
their output format.  An example of the MATLAB debugger operating
within URDB is presented in Figure~\ref{fig:matlab}.

URDB takes advantage of DMTCP's capability for checkpointing
multi-threaded processed and multiple distributed processes.
If the debugger is debugging multiple threads, then it is assumed that
the use of URDB is restricted to sending debug commands to only one of
the threads.  Similarly, for distributed computations, it is assumed
that the debugger is attached only to one of the multiple processes of
the computation.

In the case of multi-threaded programs, race conditions may alter the
execution path of a process.  For traditional debugger targets, a user
must be aware of the potential for non-determinism caused by race
conditions.  URDB does not pretend to solve this, although the future
integration of other techniques for enforcing
determinism~\cite{KendoMultithreading09,DMP09} may alleviate this
problem.

URDB also uses DMTCP's support for checkpointing ptrace-based applications.
Gdb uses ptrace, while MATLAB, python and perl do not.

\subsection{Algorithm}
\label{sec:algorithm}

The debug monitor sits between the end user and the target debugger.
For the most part, it passes user commands to the debugger and returns
the debugger output (including interrupts (\^C) by the user).  Certain
URDB commands are intercepted by the debug monitor, whereupon the
debug monitor converses with both the target debugger and the DMTCP
checkpoint-restart system.  The target debugger is assumed to support
four standard debugger commands: {\tt next} (do not step into any
function calls); {\tt step} (step into a function call); {\tt
  breakpoint} (set a breakpoint); and {\tt continue} (until next
breakpoint).  Where the debugger also provides {\tt finish} (until end
of function), URDB will support the use of that command by the end
user.  The history of executions of these commands is recorded in a
user command history.

Commands supported by the debug monitor include: {\tt checkpoint},
{\tt restore}, {\tt undo-command}, {\tt reverse-next}, {\tt
  reverse-step}, {\tt reverse-continue}, {\tt reverse-finish}, and
{\tt reverse-watch}.  Undo-command undoes the last command to the
debugger.  Reverse-next (or reverse-step) takes the process backwards
in time to the earliest statement such that a ``next'' debugger command
(or ``step'' debugger command) would return the process to the current state.
(If the process is at the first statement of a function, then
reverse-next returns the process back to the statement that called
the current function.)  Reverse-finish returns the process to the
statement of the function that called the current function.
For ease of exposition, we speak of statements, but in fact,
a {\em statement event} is intended, the last time prior to the
present when the indicated statement was executed.

The last command, {\tt reverse-watch}, uses binary search on expressions.
If a process
executes $n$~statements, then binary search will evaluate the expression
at most $\log_2 n$ times.  
This is intended as a more general variation of data watchpoints, since
the expression may depend on many addresses in RAM.   While some standard
debuggers also provide expression watchpoints, the URDB approach
is more efficient, since standard debuggers must
evaluate the expression before each of the statements.
See Section~\ref{sec:binarySearch} for a fuller
description.

The debug monitor spontaneously generates additional checkpoint images (not
necessarily requested by the user) when the time since the last
checkpoint grows beyond a specified threshold.

\paragraph{Timeout for ``continue''.}
Since the command {\tt continue} followed by {\tt reverse-step} is
problematic if the ``continue'' command had executed for a long time,
URDB places a timeout on ``continue''.  If ``continue'' executes
beyond the timeout interval, then URDB tells the target debugger to
stop execution.  URDB then creates an additional checkpoint, and then
URDB issues an additional ``continue'' command.

The timeout interval is chosen so as to maintain a reasonable user
response time for URDB.  The ``continue'' command executes natively in
many debuggers, while the ``step'' command is interpreted.  Hence the
specified timeout interval guarantees that if a single ``continue''
command is expanded into a series of ``step'' commands, then the
resulting interpreted execution (step-by-step) will complete in
reasonable time.  This turns out to be important to guarantee
reasonable response times for the algorithms below.

\paragraph{Algorithmic notation.}
The algorithm can be expressed more clearly using a modified rewrite
rule notation.  A {\em user command history} is available at each
moment when the target debugger is stopped.  The command history is
maintained for three target debugger commands: ``step'', ``next'', and
``continue'' (and optionally ``finish'').  The commands are notated by
the four characters, {\tt s}, {\tt n}, {\tt c}, and {\tt f},
respectively.  A regular expression-like notation is used for command
histories.  An asterisk, ``{\tt *}'', denotes 0 or more occurrences of
a target debugger command.  A question mark, ``{\tt ?}'', denotes
exactly one target debugger command.  A period, ``{\tt .}''
is used to syntactically separate tokens and has no further significance.
Hence, a command history {\tt *.c.n} denotes an arbitrary sequence of
commands followed by {\tt continue} and {\tt next}.

\def\ra{$\rightarrow$}

A rewrite rule, {\tt *.n \ra *}, denotes to match a command history
for which the last command was {\tt next}.  If a match is found,
then truncate the final ``next'' command from the command history,
and use the checkpoint-restart facility of DMTCP to re-execute the
debugging session from the last checkpoint until just before the
``next'' command was issued.  Similarly, a rule {\tt * \ra *.c}
says to simply execute the ``continue'' command in the target debugger.
Thus, the URDB command {\tt undo-command} would be algorithmically
described by {\tt *.? \ra *}.

\paragraph{Process state.}
At any given time, the target application being debugged is in a
unique {\em process state}.
In particular, there is a distinguished process state, {\tt
  ORIG\_STATE}, representing the state of the process at the
time that a URDB command is issued.  As rewrite rules are executed,
the process will travel backward and forward in time, and the command
history will be adjusted accordingly.  The URDB
algorithm never travels beyond {\tt ORIG\_STATE} (see
Section~\ref{sec:monitorImplementation}).

As the rewrite rules execute, the process will have a current state
will be exactly one of three states: {\tt DEEPER}, {\tt SHALLOWER},
{\tt SAME}.  This is important, since the current state will determine
which set of rewrite rules should be applied.
If the stack depth (number of call frames on the stack)
at the current time is greater than the original stack depth (depth at
the time of {\tt ORIG\_STATE}), then the current state is denoted {\tt
  DEEPER}.  Similarly, if the stack depth is less than original stack
depth, then the current state is {\tt SHALLOWER}.  Finally, if the stack
depths are equal, then the current state is {\tt SAME}.

The state {\tt SAME} is divided into two substates: {\tt ORIG\_STATE}
(which we saw above), and \\
{\tt NOT\_ORIG\_STATE}.  Since the algorithm
never travels forward in time beyond \\ {\tt ORIG\_STATE}, {\tt
  NOT\_ORIG\_STATE} always represents a prior time in the process
history.  See Section~\ref{sec:monitorImplementation} for a
description of how {\tt ORIG\_STATE} is detected.

\paragraph{The algorithms.}

The algorithm begins execution in state {\tt SAME/ORIG\_STATE}.
Rewrite rules from the current state are executed until a ``Return''
statement is executed.  If two rewrite rules within a given state both
match the current user command history, then only the first of the two
rewrite rules is applied.  The notation {\tt STATE(* $\rightarrow$ *.n)}
denotes the state that would result from executing the indicated
rewrite rule.  If a rewrite rule contains an {\em action} in square brackets,
then the action is executed after the rewrite rule is executed.
A function {\tt LEVEL(*.n)} indicates the stack depth of the user command
history from the corresponding rewrite rule.
\medskip

\def\IN{\hbox{\ \ }}
{\tt
\noindent
Algorithm REVERSE\_NEXT ( ORIG\_STATE ) : \\
\IN case DEEPER: \\
\IN\IN If STATE(* $\rightarrow$ *.n) = ORIG\_STATE, then: \\
\IN\IN\IN Call Algorithm BACK\_UP\_TO\_SAME() \\
\IN\IN * $\rightarrow$ *.n \\
\IN case SHALLOWER: \\
\IN\IN * $\rightarrow$ *.s \\
\IN case SAME: \\
\IN\IN case ORIG\_STATE: \\
\IN\IN\IN *.n $\rightarrow$ * [ If LEVEL(*.n) = LEVEL(*), then Return ] \\
\IN\IN\IN *.s $\rightarrow$ * [ If LEVEL(*.s) >= LEVEL(*), then Return ] \\
\IN\IN\IN *.? $\rightarrow$ * \\
\IN\IN case NOT\_ORIG\_STATE: \\
\IN\IN\IN * $\rightarrow$ *.n \\
\\
\noindent
Algorithm BACK\_UP\_TO\_SAME () : \\
\IN case DEEPER: \\
\IN\IN *.? $\rightarrow$ * \\
\IN case SHALLOWER: \\
\IN\IN * $\rightarrow$ *.c \\
\IN case SAME: \\
\IN\IN Return \\
\\
\noindent
Algorithm REVERSE\_STEP ( ORIG\_STATE ) : \\
\IN case DEEPER: \\
\IN\IN If STATE(* $\rightarrow$ *.n) = ORIG\_STATE, then: \\
\IN\IN\IN Call Algorithm BACK\_UP\_TO\_SAME() \\
\IN\IN * $\rightarrow$ *.n \\
\IN case SHALLOWER: \\
\IN\IN * $\rightarrow$ *.s \\
\IN case SAME: \\
\IN\IN case ORIG\_STATE \\
\IN\IN\IN *.s $\rightarrow$ * ; then Return \\
\IN\IN\IN *.? $\rightarrow$ * \\
\IN\IN case NOT\_ORIG\_STATE \\
\IN\IN\IN If STATE(* $\rightarrow$ *.s) = ORIG\_STATE, then: \\
\IN\IN\IN\IN * $\rightarrow$ *.s \\
\IN\IN\IN Else if STATE(* $\rightarrow$ *.n) = ORIG\_STATE, then: \\
\IN\IN\IN\IN * $\rightarrow$ *.s \\
\IN\IN\IN Else: \\
\IN\IN\IN\IN * $\rightarrow$ *.n \\
}

The algorithms {\tt REVERSE\_CONTINUE} and {\tt REVERSE\_FINISH} are presented
in higher level pseudo-code.

{\tt
\noindent
REVERSE\_CONTINUE ( ) : \\
\IN BEGIN: Restore last checkpoint \\
\IN\IN Repeat continue until ORIG\_STATE \\
\IN\IN if more than one previous breakpoint found \\
\IN\IN\IN Restore last checkpoint and re-execute until following breakpoint \\
\IN\IN If no previous breakpoint found, \\
\IN\IN\IN set ORIG\_STATE to state (time) as of last checkpoint \\
\IN\IN\IN Restore checkpoint previous to last checkpoint \\
\IN\IN\IN GOTO BEGIN \\
}

Here, {\tt REVERSE\_FINISH} is defined as going backward in time
to the site at which the current function was called.

\medskip
{\tt
\noindent
REVERSE\_FINISH ( ORIG\_STATE ): \\
\IN Repeat \\
\IN\IN REVERSE\_NEXT() \\
\IN Until reaching shallower level. \\
\IN Repeat \\
\IN\IN REVERSE\_NEXT() \\
\IN Until reaching beginning of that function \\
}

In situations where the time interval of {\tt REVERSE\_CONTINUE}
is excessively long, a binary search strategy can be used as 
for expression watchpoints.

When a user requests {\tt undo-command k} for some integer~$k$,
URDB restores the last checkpoint and then
re-executes the first $n-k$ of the user commands since  the last
checkpoint.

The {\tt reverse-continue} command is implemented in one of two ways.
If the time since the last checkpoint is small enough, then repeated
{\tt continue} commands are executed from the last checkpoint.
(A breakpoint is set at the current position so that the
checkpoint does not go beyond the current time.)
A count of the number~$n$ of {\tt continue} commands is maintained.
Upon discovering the last breakpoint, prior to the current time,
the process is re=executed with $n-1$ {\tt continue} commands.

If no breakpoints were hit between the last checkpoint and the current time,
then the checkpoint prior to that one is restored, with the ``current
time'' being set to the last checkpoint.  The {\tt reverse-continue}
algorithm is then repeated.

Finally, it can happen that the time from the last checkpoint to the
current time is above the specified threshold.  In that case,
a binary search strategy is employed (see Section~\ref{sec:binarySearch}),
with the condition of the binary search being to test whether
a breakpoint was seen.  If a breakpoint is found in both the first
and second half of a binary search, search recursively explores
the second half, in order to determine the last breakpoint encountered.
When the time interval being explored falls below a specified
threshold, the {\tt reverse-continue} algorithm falls back to the
previous strategy of repeated {\tt continue} commands.

\subsection{Temporal Search:  Reverse Expression Watchpoints via Binary Search}
\label{sec:binarySearch}

Reverse expression watchpoints are implemented as a showcase for
the method of temporal search.  Additional temporal search methods
can be added in the future.

A user executing a {\em reverse expression watchpoint} asks for the last
time that an expression had a value different from the current value.
The term {\em reverse} is used to emphasize that the expression to search
on is declared only after the program has finished executing the region
of interest.

We use the term {\em expression watchpoint} because this feature can
also be viewed as a generalization of the well-known data watchpoints
that many debuggers.  Current debuggers are either limited to data
watchpoints (detecting a change of contents at a specified memory
address) in order to take advantage of virtual memory hardware
support; or else they support expression watchpoints inefficiently by
single-stepping through a program and testing at each step if the
value of the given expression has change.  Unlike current debuggers,
URDB supports expression watchpoints running natively at the full
speed of the CPU.

Hence, if a program is run for $n$~steps, then traditional expression
watchpoints require $n$~executions of the target expression.  For
large~$n$, this is inefficient.  The binary search algorithm presented
here uses only $O(\log n)$ executions of debugger commands ({\tt step},
{\tt next}, and {\tt continue}) along with $O(\log n)$ evaluations of the
expression, and $O(\log n)$ checkpoints and restart.  Hence for moderate
or large~$n$, the extra cost of checkpoint and restart is justified by
the savings in not having to execute the target expression $n$~times.
The linked list example of Section~\ref{sec:introduction} presents
an example where expression watchpoints are desirable.

URDB has two modes of operation for binary search:  command history-based
binary search and time-based binary search.  In {\em time-based binary
search}, the binary search executes for half of the total time duration
between the start point and the process state at which the reverse
expression watchpoint command was given.  Once the time interval of the
binary search is reduced below a specified threshold, URDB switches
to a command history-based binary search.  (As of this writing,
time-based binary search is still being implemented.)

In {\em command history-based binary} search, we take advantage of
a natural hierarchy of debugging commands.  In URDB, this hierarchy
consists of commands to:
\begin{enumerate}
  \item restart from the next checkpoint image.
  \item issue a ``continue'' command until the next breakpoint.
  \item issue a ``next'' command that {\em steps over} any function calls.
  \item issue a ``step'' command that {\em steps into} any function calls.
\end{enumerate}
Recall that the ``continue'' and ``next'' debugger commands are
always limited in time.
This is because URDB interrupts the computation if a debugger command
has not completed within a specified timeout.  It then
creates an additional checkpoint image.
For ``continue'', an additional ``continue'' command is issued after
creating the checkpoint image.  For ``next'', URDB inserts a
temporary breakpoint at the statement following the statement at
which ``next'' was issued.  Then, an additional ``continue'' command
is issued, as in the previous case.  (This use of timeouts is still
under implementation.)

Hence, any command history can always be converted into a relatively short
sequence of commands by replacing some commands by a command to restart
from the next checkpoint image.  Binary search then uses the
sequence of debugger commands in the user history as  the process ``time''.
Through binary search, URDB determines a prefix of the user history
such that the expression under consideration has a different
value from the current one after executing the prefix of the history,
but the expression would have the same value upon executing
the next debugger command of the history.  (Note that such a prefix is not
necessarily unique, depending on the expression chosen by the user.)

Next, if the debugger command immediately following the prefix
is ``go to next checkpoint image'', then it is replaced by repeated
``next'' statements, and binary search is performed on that sequence.
If it is ``continue'', it is replaced by MAX\_SEQ
repeated ``next'' commands, followed by a ``continue'', and binary
search is performed on that sequence.
If it is ``next'', it is replaced by ``step'' followed by MAX\_SEQ
repeated ``next'' commands, and binary
search is performed on that sequence.
If it is ``step'', then we are guaranteed to have found the statement
satisfying our temporal search.

Since the time between checkpoints is limited, the number of
repeated ``next'' commands is limited.  But it may still be very large.
So, the value of MAX\_SEQ must be tuned.  One possible heuristic is
to set it initially to the length of the current history (so as
to double the history size).  If there is no difference in the
value of the expression between the beginning and end of the history,
then one can double MAX\_SEQ and try again.  Hence, the number
of debugging statements executed in this phase, to determine when
the expression value changes, can be proved to never be more
than four times the optimal number of statements.  Further,
this adds at most $\log_2 n$ evaluations of the expression
where $n$~is the optimal number of statements before the expression
value changes.

Of course, the success of these heuristics depends on the user
expression being well-behaved.  For expression values that
change monotonically over the process execution, and even for some
still more general situations, the heuristics above can be proved to work
efficiently.

\subsection{Reversible Debugging of Distributed Computations}
\label{sec:distributedComputation}

The ability of DMTCP to checkpoint distributed computations allows URDB
to reversibly debug distributed computations, subject to the same caveats
as for multi-threaded debugging.  It is assumed that the user is aware
of the issues of non-determinism for distributed computations.  Hence,
after executing a reverse-next, for example, it can happen that the
process is in a different state than originally, due to race conditions.

URDB can produce this undesirable behavior in the presence of race
conditions.  Nevertheless, the capability to reversibly debug over
shorter intervals (where there may be few or no race conditions) is
potentially valuable.  An example of where this can be useful is in
debugging how one arrived at a deadlock or livelock situation.

Further, many distributed computations produce only ``local race
conditions'' that only affect program execution over a limited time
interval, while maintaining convergence toward a globally unique
solution.  This is related to the {\em output determinism} concept of
Park \hbox{et~al.}~\cite{ParkEtAl09}.  In these situations,
non-deterministic reversible debugging may also prove useful.  We are
still gaining experience from the end user perspective concerning the
benefits of non-deterministic reversible debugging.

\subsection{Implementation of the Debug Monitor}
\label{sec:monitorImplementation}

The debug monitor uses the concept of breakpoint events.  A {\em
  breakpoint event} is a process state (time) at which a breakpoint
was hit.  If one hits the same breakpoint multiple times, one must
distinguish the different breakpoint events.  In gdb and some other
debuggers, one can determine a {\em unique} breakpoint event as the
debugger travels backward and forward in time.  One does this by
noting the number of times that a breakpoint was hit by the debugger.
For example, gdb reports that number through the gdb command {\tt info
  breakpoints}.

For debuggers that do not associate a number with each breakpoint,
the filename and line number of that breakpoint serve the same purpose.
Some debuggers do not record the number of times a particular breakpoint
was hit.  In those cases, URDB can increment a variable of the target
debugger (or even a global variable in the target application) with the
total number of breakpoints hit so far.  As the process moves backward
and forward in time, this number is automatically updated.  In this case,
a breakpoint event is simply the number of breakpoint hits encountered
since the beginning of the process.

A key to the URDB algorithm is being able to recognize the {\tt
  ORIG\_STATE} of the preceding section.  The implementation defines a
process state to be a triple: (filename, line number, last breakpoint
event seen).  The current process state is considered to be in the
{\tt ORIG\_STATE} when it has the same value of the triple.  There may
be more than one process state having the same triple.  The key to the
correctness of the algorithm is that when the current process state
has the same triple as the original process state, then the two
process states must have the same command history through the last
``continue'' statement.  Hence, any potential differences in the
command history concern only sequences of ``step'' and ``next''
instructions, which are directly handled by the algorithm.

As an optimization, coalescing of debugger commands such as ``next''
and ``step'' is employed.  For example, a debugger command ``next;
next; next'' can be replaced by ``next 3'' for greater efficiency.
The issue of debugger breakpoints adds a subtle point to the
implementation.  All breakpoints must be temporarily disabled or
deleted during a repeated ``next''.  Yet breakpoints and
disable/delete commands in the user history must continue to be
faithfully re-executed.

\section{Experimental Results}
\label{sec:experiment}

Unless otherwise indicated, all experiments are done under Ubuntu~9.04
with a Linux kernel~2.6.28.  The computer used has two Intel Core~2 Duo
CPUs (four cores) running at 3.0~GHz.  The experiments used gcc~4.3.3,
gdb-6.8, python~2.6.2, perl~5.10.0 and MATLAB~7.4.0.  MATLAB was used on
a four-core Intel Xeon running at 1.86~GHz.  The version of DMTCP used
was revision~393 (unstable branch with support for ptrace) of the
DMTCP subversion repository.  URDB revision~109 of the URDB subversion
repository was used.  DMTCP was configured to not use compression.
This produces larger checkpoint images, but checkpoint and restart
are faster.

URDB consists of a python-based debug monitor (and a smaller C++ file).
The python monitor component contains 1,000 lines of code.
URDB also requires a debugger-specific python file.
For each of its four target debuggers.
The python-based personality template expands to a debugger-specific file
containing:  204~lines of code for MATLAB;
242~lines of code for python (pdb module);
162~lines of code for perl (perl~-d);
and 258~lines of code in the case of gdb.
DMTCP consisted of 22,300 lines of code.

The experiments are grouped into four parts:
\begin{enumerate}
\item tests done across all four personalities (target debuggers);
\item tests done only on the gdb personality (implementation is planned
	later for the remaining personalities);
\item scalability tests with the gdb personality; and
\item demonstrations of unusual URDB applications.
\end{enumerate}

\begin{table}[htb]
\centering
\begin{tabular}{|c|c|c|c|}
\hline
Language & reverse-next (s) & reverse-step (s) & checkpoint (s)\\
 \hline
 \hline
gdb-6.8 & 0.68 & 3.17 & 4.52\\
  \hline
MATLAB & 0.72  & 0.92 & 6.08 \\
  \hline
perl & 0.97 & 1.78 & 4.37 \\
 \hline
python & 0.29 & 2.90 & 4.33 \\
\hline
\end{tabular}
\caption{Times for reverse-next, reverse-step, checkpoint} 
\label{tab:all-debuggers}
\end{table}

\begin{table}[htb]
\centering
\begin{tabular}{|c|c|c|c|}
\hline
Language & checkpoint & restart & size of checkpoint files\\
 \hline
 \hline
gdb-6.8 & 1.08 & 0.42 & 11.08 \\
  \hline
MATLAB & 4.65 & 1.69 & 32.30 \\
  \hline
perl & 0.43 & 0.20 & 3.42 \\
 \hline
python & 0.45 & 0.19 & 3.66 \\
\hline
\end{tabular}
\caption{Times for checkpoint and restart (seconds),
	 as well as the sizes of the checkpoint image files (MB)} 
\label{tab:dmtcp-data}
\end{table}

\paragraph{Experiments across reversible debuggers.}
Timings for the URDB reversible debugger commands for each of the four
target debuggers are contained in Table~\ref{tab:all-debuggers}.
The reversible debuggers tested are for gdb~6.8, MATLAB,
python (pdb module), and perl (perl -d).  (The more recent
gdb~7.0 debugger became available only during the final phases
of the current work.)  Tests are reported
for reverse-next, reverse-step, and time of URDB
to checkpoint.  Checkpoint image sizes under URDB
are within 0.1\% of the sizes for DMTCP alone, as given
in Table~\ref{tab:dmtcp-data}.

Gdb was tested with a main function with a loop, and the body of the loop
calling a second function.  MATLAB was tested on a short program with
three functions (function~A calling function~B, which called function~C).
Python and perl were similarly tested, except with just four functions
(a main function with a loop, and the body of the loop calling each of
function~A, function~B, and function~C).

Table~\ref{tab:dmtcp-data} reports on the underlying times and storage
for DMTCP alone (without using URDB).  Times for checkpoint, restart,
and the size of the checkpoint image files are reported.  In the case of
gdb, there are two checkpoint images: gdb (9.5~MB); and the target
application a.out (1.6~MB).  MATLAB employs two checkpoint images: MATLAB
(30.3~MB); and matlab\_helper (2.0~MB).  Python and perl consist of a
single checkpoint image each of sizes~3.66 and 3.42~MB, respectively.
The sizes of the checkpoint image files are reported in 
Table~\ref{tab:dmtcp-data}.
In the case of gdb and MATLAB,
Table~\ref{tab:dmtcp-data} reports the combined sizes of the two checkpoint 
images.

\paragraph{Experiments specific to the gdb reversible debuggers.}
The following additional URDB functionality is currently gdb-specific,
but will soon be ported to all
reversible debuggers.  As of this writing, URDB also supports
reverse-continue, reverse-finish and reverse expression watchpoints.
Table~\ref{tab:urdb-gdb-only} reports the corresponding times.

\begin{table}[htb]
\centering
\begin{tabular}{|c|c|}
\hline
Command & Time \\
 \hline
 \hline
reverse-continue & 1.13 \\
  \hline
reverse-finish &  2.11 \\
  \hline
reverse expression watchpoint & 36.86 \\
 \hline
\end{tabular}
\caption{Times for reverse-continue, reverse-finish and reverse expression 
watchpoint (seconds) } 
\label{tab:urdb-gdb-only}
\end{table}

\paragraph{Experiments for scalability.}

\begin{table}[htb]
\centering
\begin{tabular}{|c|c|c|c|}
\hline
Command & After 10 & After 100 & After 1000 \\
        & commands & commands & commands \\
 \hline
 \hline
reverse-next & 0.57 &  0.69  & 0.79 \\
  \hline
reverse-step & 0.75 & 0.94  & 0.87\\
  \hline
\end{tabular}
\caption{Times for reverse-next and reverse-step after 10/100/1000
  ``next'' and ``step'' commands, respectively.  All times
  are in seconds. }
\label{tab:scalability}
\end{table}

The scalability experiments were performed using gdb-6.8. 
Table~\ref{tab:scalability} reports on scalability trends.
Similar trends are expected for the other reversible debuggers.
The primary programs
for testing were a recursive version of factorial and an iterative
version of factorial.  Both versions use floating point arguments
instead of integers in order to be able to run on longer computations.
The base case tests if the argument~$x$ is between 0.5 and~1.5 and then
returns~1.0 for factorial($x$) for such~$x$.  Reverse-next is
tested on the iterative version of factorial for the sake of the
long execution sequence within a single function.
Reverse-continue implicitly requires function calls, and hence
is tested on the recursive version of factorial.

In each case, a checkpoint was previously taken at the beginning of
main, and the program is run for the indicated number of statements.
The dominating time is the time for the checkpoint.  Hence, the
decreased time for the largest number of ``step'' commands is due to
the variation in times for restarting from the checkpoint.  This is
because these examples use coalescing of debugger commands (see
Section~{sec:monitorImplementation}).  In the case of reverse-step,
every two debugger ``step'' commands cause the application to
enter a new call frame.
The corresponding times for the same experiment, with coalescing
turned off, are shown in Table~\ref{tab:no-coal}.

\begin{table}[htb]
\centering
\begin{tabular}{|c|c|c|c|}
\hline
Command & After 10 & After 100 & After 1000 \\
        & commands & commands & commands \\
 \hline
 \hline
reverse-next & 0.80  & 4.57 & 40.75 \\
  \hline
reverse-step & 1.06 & 4.75 & 41.13 \\
  \hline
\end{tabular}
\caption{Times for reverse-next and reverse-step after 10/100/1000
  ``next'' and ``step'' commands, respectively, without using 
coalescing.  All times are in seconds. }
\label{tab:no-coal}
\end{table}

\paragraph{Demonstrations of unusual URDB applications.}
URDB was also tested on two more unusual applications in order
to demonstrate the generality of URDB.  The two examples are:
\begin{enumerate}
  \item {\em interactive simulations and games:}  Two classic 1980-era
	BSD UNIX curses-based games, robots and rogue, were chosen
	to demonstrate this category.
	(See the Wikipedia articles ``Rogue (computer game)'' and
	``Robots (computer game)''.)
	A modified URDB was employed with control-E/control-T for
	checkpoint/undo, respectively, since interactive games do
	not read commands through standard input.
  \item {\em debugging multiple processes:}  the example of gdb debugging
	a second gdb process that was itself debugging an application
	a.out process.  This example is interesting
	since the first gdb is using ptrace to trace the second gdb,
	while the second gdb is similarly using ptrace to trace
	the a.out application.
\end{enumerate}

The version of
``robots'' used was version 2.16 of the bsdgames Ubuntu package,
while ``rogue'' was version 2.17 of the bsdgames-nonfree Ubuntu package.
The times for checkpoint and undo for robots were 3.67 and~ 0.04 seconds,
respectively.
Similarly, the times for checkpoint and undo for rogue
were 3.21 and~0.05, respectively.  The checkpoint images were
of size 12.5 and 12.8~MB for robots and rogue, respectively.

\begin{table}[htb]
\centering
\begin{tabular}{|c|c|}
\hline
Command & Times \\
 \hline
 \hline
reverse-next & 0.90 \\
  \hline
reverse-step & 0.94 \\
  \hline
checkpoint & 5.17 \\
  \hline
\end{tabular}
\caption{Times for reverse-next, reverse-step and checkpoint for 
gdb debugging gdb debugging a.out (seconds) } 
\label{tab:gdb-debugging-gdb}
\end{table}

Times and sizes for reverse-next, reverse-step, and checkpoint are
reported for gdb debugging a second gdb that in turn was 
debugging~a.out in Table~\ref{tab:gdb-debugging-gdb}.

\subsection{Future Directions:  Meta-Programming on Process Lifetimes}

We view the demonstration of universal reversible debugging not as an end
in itself, but rather as a beginning with multiple other possibilities.
The URDB approach demonstrates the possibility of meta-computing over
process lifetimes for arbitrary processes.  Expression watchpoints (see
Section~\ref{sec:binarySearch}) provide an example of meta-computing,
in which we execute binary search over a process lifetime.

Eventually, URDB will be extended into a meta-computing language that
treats processes as {\em first class objects}.  Files can be replicated,
copied to new hosts, be modified, and support seeks to arbitrary locations
in the file.  The same will be true of can processes.

In particular, {\em speculative execution of programs} and {\em program
introspection} becomes possible under the control of the program itself.
As an example of the former possibility, programs can speculatively
execute themselves and avoid crashes, deadlocks and livelocks.  As an
example of the latter possibility, upon finding a solution to a problem,
a program can use program introspection to inspect its execution history
and produce a short explanation of steps needs to produce the solution.

Another novel use is comparison of two versions of the same program.
One can request, for example, to ignore all call frames except the first
four.  One then begins execution of the two programs and asks for the
first statement for which some user-specified expression differs for
the two programs.

Building on this capability, one can imagine producing a dataflow
description (or dependency graph) of variable occurrences on which a
later variable occurrence depends.  While this is normally a difficult
task, speculative execution allows one to heuristically test whether a
dependency exists between an original variable occurrence and a later
variable occurrence.  One heuristically tests by varying the value of
the original variable occurrence and determining if the later variable
occurrence changes its value.  For this purpose, a segmentation fault
before obtaining the value of the later variable occurrence is considered
to be a change of its value.

Interactive simulations also benefit.  Assume the program execution
depends on inputs from a human user and is otherwise deterministic.
Hence, the program execution uniquely depends on an input trace.
One specifies a figure of merit for a simulation (similarly to the use
of expressions in expression watchpoints).  One can take
checkpoints, and represent a process state by the most recent checkpoint,
followed by an input trace.  Thus, the storage cost of the checkpoint
is amortized over many potential process states.  (The cost can be still
further reduced through the use of incremental checkpointing.)
This representation allows one to then employ
	best-first search, A* search,
	branch-and-bound, multiple simulation annealing runs, and
other such classical search techniques.  The enabling technology
is the ability to treat processes (or process states here)
as first class objects.

A similar idea can be employed for capturing broad test coverage
in software testing.  One employs a similarity measure on process states.
One then employs search techniques to produce all possible ``unique process
states'', where the similarity measure determines if two process
states are to be considered the same process state for this purpose.
The similarity measure can be applied to the process especially easily
if the process is running under a debugger that has access to its
internal variables.

\section{Conclusion and Future Work}
\label{sec:conclusion}

Incremental checkpointing is planned for a future version of URDB in
order both to reduce the size of each checkpoint and to improve the
times for checkpointing.  The use of temporary checkpoints of debugger
commands will then be added to greatly enhance the speed of expression
watchpoints.

Experiments that mix the URDB approach with event logging are also planned.

Finally, expression watchpoints are an example of {\em program-based
introspection}: when in my process lifetime did this expression change
its value?  The demonstrations for the interactive games {\em rogue}
and {\em robots} are a form of {\em speculative program execution}.
These rudimentary examples open the way for more sophisticated
applications of: {\em time-traveling program-based introspection};
and {\em speculative program execution}.

\section{Acknowledgement}

We wish to thank Priyadarshan Vyas for his contributions to an earlier
version of the monitor able to support the {\em undo} command described
in this paper.  We also thank Shobhit Agarwal and Amruta Chougule for
their earlier identification of several issues and suggested solutions
for checkpointing gdb with DMTCP.  We also wish to thank Peter Desnoyers
and Jason Ansel for their helpful discussions and insights.

\bibliographystyle{alpha}
\bibliography{urdb-debugger}

\end{document}